\documentclass[aps,prl,twocolumn,superscriptaddress]{revtex4}

\usepackage{amsmath,amssymb,graphicx,color}

\begin{document}

\title{Single photons from coupled quantum modes}

\author{T. C. H. Liew}
\author{V. Savona}
\affiliation{Institute of Theoretical Physics, Ecole Polytechnique F\'{e}d\'{e}rale de Lausanne (EPFL), CH-1015 Lausanne, Switzerland}

\date{\today}

\begin{abstract}
Single photon emitters often rely on a strong nonlinearity to make the behaviour of a quantum mode susceptible to a change in the number of quanta between one and two. In most systems the strength of nonlinearity is weak, such that changes at the single quantum level have little effect. Here, we consider coupled quantum modes and find that they can be strongly sensitive at the single quantum level, even if nonlinear interactions are modest. As examples, we consider solid-state implementations based on the tunneling of polaritons between quantum boxes or their parametric modes in a microcavity. We find that these systems can act as promising single photon emitters.
\end{abstract}

\maketitle

{\it Introduction.}--- The construction of single photon sources~\cite{Lounis2005,Shields2007} is a current aim of quantum nonlinear optics. Aside contributing to the security of quantum cryptography~\cite{Scarani2009}, single photon sources are useful elsewhere, for example in schemes for quantum computation using only linear optics and photodetection~\cite{Knill2001}. For some applications it is enough to reduce the intensity of a laser source to obtain single photons with a probability limited by the Poisson distribution. To do better than Poisson statistics one requires some form of nonlinearity. However, when one works in the single photon regime a strong nonlinearity is not so easy to find.

In semiconductor microcavities, light is strongly coupled to quantum well excitons resulting in new quasiparticles known as polaritons. Taking the best from both parents, polaritons have attracted particular attention for over a decade due to their strong nonlinearity (inherited from excitons) as well as their fast dynamics, long coherence and ability to couple to external light (features of photons). Polariton-polariton interactions have resulted in micron-sized optical parametric oscillators~\cite{Tartakovskii2000,Stevenson2000,Diederichs2006}, optical gates~\cite{Leyder2007}, spontaneous coherence~\cite{Kasprzak2006,Balili2007,Lai2007}, low threshold lasing at room temperature~\cite{Imamoglu1996,Christopoulos2007,Bajoni2008} and superfluidity~\cite{Amo2009}. Whilst these effects involve many polaritons at once, we wish to focus on the single quantum regime. In planar cavities, quantum effects such as squeezing have been reported and several studies on quantum correlations undertaken~\cite{Romanelli2005,Karr2004,Savasta2005,Verger2007}. More pronounced effects at the single polariton level are expected in quantum boxes~\cite{Bloch1997,IdrissiKaitouni2006,Bajoni2007}, where polaritons are fully confined in three-dimensions and forced to interact even more strongly. Available recently, such confinement has encouraging prospects for single photon sources.

It has been predicted that for a very strong nonlinearity, the presence of a single polariton can block the resonant injection of another~\cite{Verger2006}, analogous to the photon blockade~\cite{Imamoglu1997} of nonlinear cavities. However, to obtain a strong enough nonlinearity for a single photon source, an extremely small quantum box is required (with size of the order of $200$nm). Although one may anticipate such a system in the future, current systems do not display such a strong nonlinearity - whilst high nonlinearity is present in semiconductor microcavities, the energy shift caused by two interacting polaritons remains small.

We consider theoretically two coupled quantum boxes and show that the coupling can dramatically enhance the characteristics of single photon devices. By solving the quantum master equation for the density matrix, we find strong single photon statistics for values of the polariton-polariton interaction strength corresponding to today's systems. We show that this is due to correlations between the quantum fluctuations in the two boxes, allowing a much stronger sensitivity of the system to the population compared to the single mode case. We expect that the coupling under study can also be exploited in analogous systems such as coupled nonlinear cavities or coupled photonic crystal cavities~\cite{Gerace2009}. Finally, we show how mode coupling in parametric oscillators can also enhance single photon statistics due to selection rules.

{\it A pair of linearly coupled modes.}--- Consider a pair of quantum modes described by creation operators $\hat{a}_1^\dagger$ and $\hat{a}_2^\dagger$ respectively. As an example, we imagine the lowest energy polariton modes of two spatially separated microcavity quantum boxes. In each box, polariton-polariton interactions are characterized by an interaction strength $\alpha$. The boxes are spatially separated such that there are no significant nonlinear interactions between boxes. However, the boxes are close enough together such that particles can tunnel from one box to the other, at a rate given by the tunneling constant $J$. The Hamiltonian is:
\begin{align}
\hat{\mathcal{H}}&=E_1\hat{a}^\dagger_1\hat{a}_1+E_2\hat{a}^\dagger_2\hat{a}_2+\alpha\left(\hat{a}^\dagger_1\hat{a}^\dagger_1\hat{a}_1\hat{a}_1+\hat{a}^\dagger_2\hat{a}^\dagger_2\hat{a}_2\hat{a}_2\right)\notag\\
&\hspace{5mm}-J(\hat{a}^\dagger_1\hat{a}_2+\hat{a}^\dagger_2\hat{a}_1)+F\hat{a}^\dagger_1+F^*\hat{a}_1\label{eq:2modeHam}
\end{align}
where $E_1$ and $E_2$ are the uncoupled energy levels of the two quantum modes and $F$ represents a coherent excitation of the first mode. With quantum boxes, this would be a laser excitation focused onto the first quantum box. We define our energy scale such that the pump energy is zero. The evolution equation of the corresponding density matrix, $\boldsymbol{\rho}$, is:
\begin{equation}
i\hbar\frac{d \boldsymbol{\rho}}{dt}=\left[\hat{\mathcal{H}},\boldsymbol{\rho}\right]+i\frac{\Gamma}{2}\sum_{n=1}^2\left(2\hat{a}_n\boldsymbol{\rho}\hat{a}^\dagger_n-\hat{a}^\dagger_n\hat{a}_n\boldsymbol{\rho}-\boldsymbol{\rho}\hat{a}^\dagger_n\hat{a}_n\right)\label{eq:master}
\end{equation}
where the last term represents the standard Lindblad dissipation characterized by decay rate $\Gamma$. Equation (\ref{eq:master}) can be solved by expanding the density matrix over a particle number basis in a similar way to that done in Ref.~\cite{Verger2006}; one truncates at a given particle number and propagates in time from the vacuum to the steady state~\cite{EPAPS}.

For a pair of quantum boxes of $3\mu m$ size, separated by $1\mu m$, a typical value of the tunnel constant is $J=0.5$meV. Since $J>\Gamma$, strong coupling takes place and the single particle eigenmodes are the symmetric and antisymmetric modes spanning the two wells~\cite{Sarchi2008}. We take $\alpha=0.012$meV, a value measured in Ref. ~\cite{Kasprzak2007} for condensed polaritons occupying spot sizes of $\sim3\mu m$. Although the pump acts directly on the first well it effectively pumps the second well due to an interference effect. This can be understood by considering the Heisenberg equations for the symmetric ($\hat{a}_+=\sin{\phi}\hat{a}_1+\cos{\phi}\hat{a}_2$) and antisymmetric ($\hat{a}_-=\cos{\phi}\hat{a}_1-\sin{\phi}\hat{a}_2$) field operators, with eigenenergies $E_+$ and $E_-$ respectively:
\begin{equation}
\frac{d\hat{a}_+}{dt}=E_+\hat{a}_++F\sin{\phi};\hspace{5mm}\frac{d\hat{a}_-}{dt}=E_-\hat{a}_-+F\cos{\phi}\notag
\end{equation}
For $E_1\approx E_2$, $\phi\approx\pi/4$ and in the steady state one finds that $\hat{a}_\pm=\frac{F}{E_\pm\sqrt{2}}$. Since $E_+$ and $E_-$ have different signs, $\hat{a}_+$ and $\hat{a}_-$ are excited with different signs such that $\hat{a}_2$ is excited instead of $\hat{a}_1$.

A key quantity in quantum optics is the second order correlation function, defined as:
\begin{equation}
g_{2,nm}(t-t^\prime)=\frac{\left<\hat{a}^\dagger_n(t^\prime)\hat{a}^\dagger_m(t)\hat{a}_m(t)\hat{a}_n(t^\prime)\right>}{\left<\hat{a}^\dagger_n(t^\prime)\hat{a}_n(t^\prime)\right>\left<\hat{a}^\dagger_m(t)\hat{a}_m(t)\right>}
\end{equation}
When $n$ and $m$ correspond to the same mode and $t=t^\prime$, this quantity also measures the performance of a single photon source; an ideal source has $g_{2,nn}(0)=0$, whilst a classical source has $g_{2,nn}(0)=1$.

\begin{figure}[h]
\centering
\includegraphics[width=8.116cm]{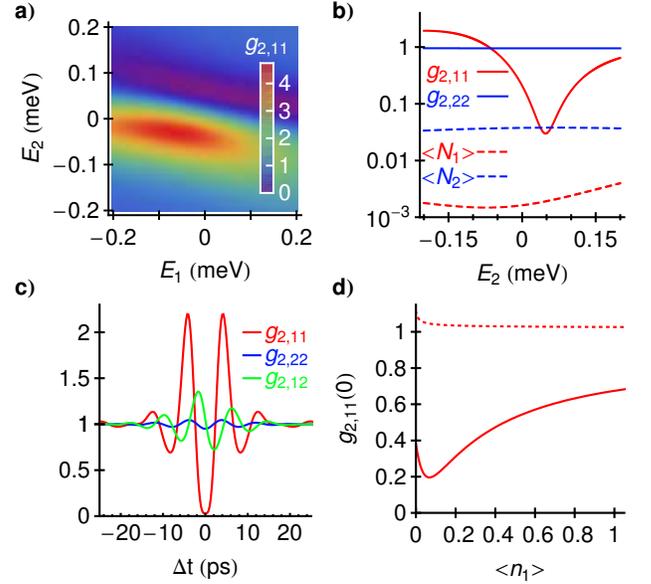}
\caption{a) Variation of $g_{2,11}(0)$ with $E_1$ and $E_2$. b) Dependence of the equal time correlation function and average populations on $E_2$ for $E_1=0.07$meV.  c) $g_{2,11}(t)$ for the optimum parameters from (a). d) $g_{2,11}(0)$ from Eq.(\ref{eq:g2Analytic}) (solid) and comparison to the single mode case (dotted). In all panels $\Gamma=0.2$meV and in (a-c) $F=0.1$meV.}
\label{fig:2modes}
\end{figure}

For fixed pump intensity, the dependence of $g_{2,11}(0)$ on the energy levels of the two wells is shown in Fig.~\ref{fig:2modes}a. The optimum (smallest) $g_{2,11}(0)$ was attained when $E_1=0.07$meV and $E_2=0.05$meV. The $g_{2,11}(0)$ depends mostly on the energy of the second well, $E_2$, and this variation is shown again in Fig.~\ref{fig:2modes}b along with the variation of $g_{2,22}(0)$, and the average well populations, $\left<N_1\right>$ and $\left<N_2\right>$ respectively. Unlike in Fig.~\ref{fig:2modes}a, the pump energy is also varied to maintain a constant detuning between the pump and the lowest energy (symmetric) single particle eigenstate. This allows a better test of the variation of $E_2$ since the average populations do not change drastically and reveals that whilst the second well has $g_{2,22}(0)\approx1$, varying $E_2$ has a dramatic effect on $g_{2,11}(0)$.

Utilizing the first well as a single photon source, one finds that at a pump amplitude giving $\left<N_1\right>=0.02$ the probability of having more than one photon is $0.18\%$; this is five times better than the failure rate of devices based on spontaneous parametric down conversion~\cite{Lounis2005}. Spectral filtering could further reduce emission from the $n_1\geq2$ states, providing extra improvement. Fig.~\ref{fig:2modes}c shows the unequal time second order correlation functions~\cite{Verger2006}, which oscillate at half the Rabi oscillation period arising from the $J=0.5$meV coupling.

To better understand the low value of $g_{2,11}(0)$ we carried out analytical calculations, extending the method of Ref.~\cite{Carusotto2001}, which applies directly to the single mode case. We use stochastic (Langevin-type) equations for the evolution of quantum fields~\cite{Drummond1980}. Lowest-order fluctuations of the fields around their mean values can be found by solving the linearized ($\hat{a}_n\mapsto\bar{a}_n+\delta a_n$) version of the equations~\cite{EPAPS}. Choosing the convention that $\bar{a}_1$ is real, the $2^{nd}$ order correlation can be written~\cite{Drummond1980}:
\begin{equation}
g_{2,11}=1+\frac{2}{n_1}\left[\left<\delta a_1^*\delta a_1\right>+\Re e\left\{\left<\delta a_1\delta a_1\right>\right\}\right]
\end{equation}
which yields~\cite{EPAPS}:
\begin{align}
g&_{2,11}=1+\frac{2}{n_1}\left[\left<\delta\alpha_1^*\delta\alpha_1\right>\right]-\frac{2\alpha}{\left(E_1+4\alpha n_1\right)\left(1+\zeta^2\right)}\notag\\
&+\frac{2J\Re e\left\{\left<\delta a_1 \delta a_2\right>\right\}}{n_1\left(E_1+4\alpha n_1\right)\left(1+\zeta^2\right)}-\frac{2\zeta J \Im m\left\{\left<\delta a_1 \delta a_2\right>\right\}}{n_1\left(E_1+4\alpha n_1\right)\left(1+\zeta^2\right)}\label{eq:g2Analytic}
\end{align}
where $\zeta=\Gamma/\left(2(E_1+4\alpha n_1)\right)$. The mean field values, $\bar{a}_1$, can be obtained as in Ref.~\cite{Sarchi2008}. To calculate all second order correlations we have extended the method of Ref.~\cite{Chaturvedi1977}. In Fig.~\ref{fig:2modes}d we compare the result of Eq.~\ref{eq:g2Analytic} for the two mode (solid line) and single mode (dotted line) cases. The single mode value is obtained by setting $J=0$ and matches the result from Ref.~\cite{Carusotto2001}. For parameters corresponding to the optimum $g_{2,11}$, the last term in Eq.~\ref{eq:g2Analytic} makes a strong negative contribution~\cite{EPAPS}. In other words, the correlated noise fluctuations $\left<\delta a_1 \delta a_2\right>$ drive the low value of $g_{2,11}$. These correlations are a result of the interplay between nonlinearity (in the limit $\alpha\mapsto0$, $\left<\delta a_1 \delta a_2\right>\mapsto0$) and tunnelling. This is a very different mechanism from the polariton blockade~\cite{Carusotto2001}, which originate from the third term in Eq.~\ref{eq:g2Analytic} that vanishes in the present regime where $\alpha n_{1,2}\ll\Gamma$.

For pioneering experiments one may also consider replacing the coupled quantum box modes by the circularly polarized spin modes of a single quantum box or a localized state in a planar microcavity. Magnetic fields parallel and perpendicular to the growth direction would allow the tuning of the spin energy levels (as suggested as a control method of the original polariton blockade~\cite{Zhang2009}) and the coupling constant $J$, respectively.

\begin{figure}[h]
\centering
\includegraphics[width=8.116cm]{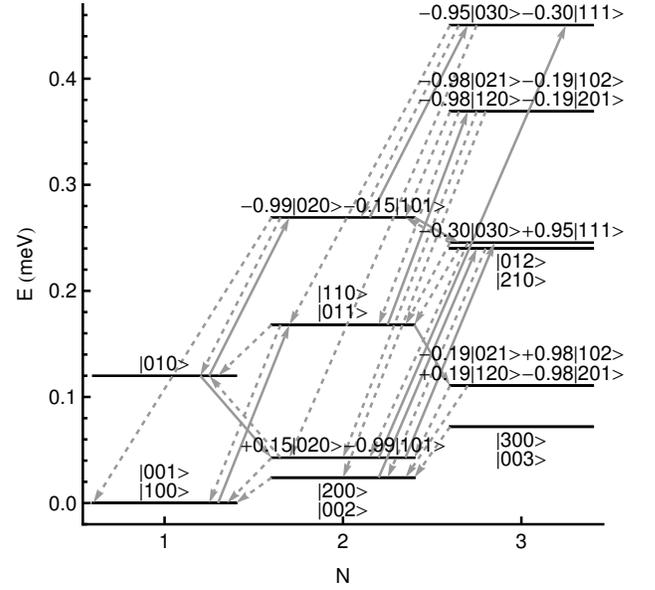}
\caption{Eigenstates from the diagonalization of the Hamiltonian Eq.(\ref{eq:3modeHam}) without the pump terms on the particle number manifold. Parameters: $E_1=0$meV, $E_1=0.12$meV, $E_2=0$meV, $\alpha=0.012$meV, $\Gamma=0.1$meV. The transitions caused by optical pumping and decay are shown by the solid and dashed arrows respectively.}
\label{fig:3modeLevels}
\end{figure}

{\it Parametrically coupled modes.}--- Given the attention devoted to parametric processes in microcavities at the beginning of the millennium~\cite{Tartakovskii2000,Stevenson2000,Diederichs2006}, it is only natural for us to ask whether sub-Poisson statistics can also be derived from pair scattering processes. Again, a variety of systems can be imagined, including the eigenmodes of a localized quantum box~\cite{Cerna2009}. The Hamiltonian now includes three separate modes characterized by creation operators $\hat{a}_1^\dagger$, $\hat{a}_2^\dagger$ and $\hat{a}_3^\dagger$ respectively:
\begin{align}
\hat{\mathcal{H}}&=\sum_{n=1}^3\left(E_n\hat{a}^\dagger_n\hat{a}_n+\alpha\hat{a}^\dagger_n\hat{a}^\dagger_n\hat{a}_n\hat{a}_n+2\alpha\sum_{m\neq n}\hat{a}^\dagger_n\hat{a}^\dagger_m\hat{a}_n\hat{a}_m\right)\notag\\
&+2\alpha\left(\hat{a}^\dagger_1\hat{a}^\dagger_3\hat{a}_2\hat{a}_2+\hat{a}^\dagger_2\hat{a}^\dagger_2\hat{a}_1\hat{a}_3\right)+F\hat{a}^\dagger_2+F^*\hat{a}_2\label{eq:3modeHam}
\end{align}
The scattering terms represent the same selection rules as when one deals with a pump mode that can scatter in pairs to signal and idler modes, as in intra-branch~\cite{Tartakovskii2000,Stevenson2000} and inter-branch~\cite{Diederichs2006} scattering in planar microcavities. The Hamiltonian can be diagonalized on a number state manifold in the absence of the pump terms. The energy levels are shown in Fig.~\ref{fig:3modeLevels}.
The only state in the $n_1=2$ manifold that can pollute the value of $g_{2,11}(0)$ is the lowest lying $\left|200\right>$ state. An analysis of the allowed transitions shows that this state can only be reached from decay of an $n_1=3$ state. In the low occupation limit the system is not expected to visit the $n_1=3$ manifold very frequently, leading us to expect a low value of $g_{2,11}(0)$.
\begin{figure}[h]
\centering
\includegraphics[width=8.116cm]{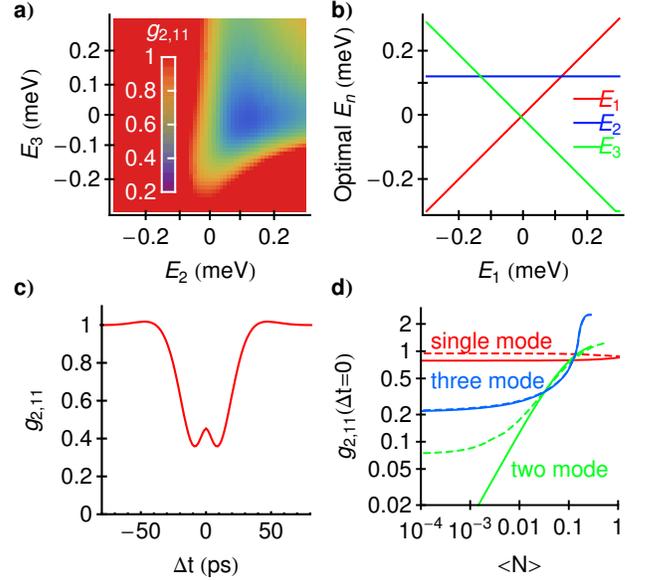}
\caption{a) Variation of $g_{2,11}(0)$ with $E_2$ and $E_3$ for $E_1=0$, $F=0.1$meV. b) The optimum energy levels. c) $g_{2,11}$ for the optimum parameters. Other parameters: $\Gamma=0.1$meV, $\alpha=0.012$meV. d) Comparison of the $g_{2,11}(0)$ available from the single mode polariton blockade (with pump resonant with the bare mode energy), the two coupled mode case (from Fig~\ref{fig:2modes}) and the three parametrically coupled modes. $g_{2,11}(0)$ is plotted as a function of the average occupation number of the signal mode. Dashed curves show the effect of dephasing.}
\label{fig:3modes}
\end{figure}

Using a similar evolution equation to Eq.(\ref{eq:master}), $g_{2,11}(0)$ is calculated in Fig.~\ref{fig:3modes}a for different values of the energy levels, $E_2$ and $E_3$. A study varying also $E_1$, showed that the optimum parameters are for the case when the pumped mode is at an energy slightly higher than the pump energy, $E_2=0.12$meV and the sum of the signal and idler mode energies is resonant with the pump (see Fig.~\ref{fig:3modes}b). In fact $g_{2,11}(0)$ is found only to vary if $E_1+E_3$ is changed, that is, a selection of values of $E_1$ and $E_3$ can be used to obtain optimum results.

The time dependent second order correlation is shown in Fig.~\ref{fig:3modes}c. For zero delay we obtain $g_{2,11}(0)=0.28$. It is important to remember that we are working with a value of the polariton-polariton interaction strength available in two dimensional planar cavities~\cite{Kasprzak2007}; much lower values of $g_{2,11}(0)$ would appear in more confined systems in which the strength of interactions is higher.

For comparison, the second order coherence function is shown in Fig.~\ref{fig:3modes}d for three cases using the same value of the interaction strength: the single mode polariton blockade~\cite{Verger2006}; the two coupled well case (studied in the previous section); and the case of three parametrically coupled modes. It is clear that the use of schemes involving two or three coupled modes can give statistics closer to that of a single photon device than the single mode case. Whilst in all schemes the value of $g_{2,11}(0)$ decreases with the average population of the signal mode, for the single mode and three mode case $g_{2,11}(0)$ tends to a constant value as $\left<N\right>$ is decreased.

Finally, we have considered the effect of pure dephasing (associated with exciton-phonon scattering) by adding the term:
\begin{equation}
\frac{\Gamma_P}{2}\sum_n\left(2\hat{a}^\dagger_n\hat{a}_n\boldsymbol{\rho}\hat{a}^\dagger_n\hat{a}_n-\hat{a}^\dagger_n\hat{a}_n\hat{a}^\dagger_n\hat{a}_n\boldsymbol{\rho}-\boldsymbol{\rho}\hat{a}^\dagger_n\hat{a}_n\hat{a}^\dagger_n\hat{a}_n\right)
\end{equation}
to the right-hand side of Eq.(\ref{eq:master}). Taking $\Gamma_P=0.3\mu eV$ (an upper estimate of the dephasing from Ref.~\cite{Savona1997}) gives the dashed curves shown in Fig.~\ref{fig:3modes}d.

{\it Conclusion.}--- We considered the use of a pair of coupled quantum boxes as a single photon source. We present the general idea that coupling can dramatically improve the single photon statistics compared to the single-mode case, through noise correlations. With competitive characteristics one may choose quantum boxes in a solid-state system, which offer fast (picosecond scale) relaxation rates; compact size; emission into a well-defined spatial mode (a luxury not always present when working with quantum dots); and wavelengths compatible with transmission through silica fibres and photo-detection with silicon based photon-counters. Alternatively, one may consider coupling nonlinear cavities or using parametrically coupled modes in planar microcavities. Indeed we anticipate more studies specific to each system and experimental verifications. The nonlinearity is available with our present technology and we have several paths to choose from.

We thank M. Wouters and A. Fiore for useful discussions. This work is supported by NCCR Quantum Photonics (NCCR QP), research instrument of the Swiss National Science Foundation (SNSF).

\end{document}